\documentclass[aps,showpacs,tightenlines,nofootinbib,prd]{revtex4}
\usepackage{amsmath,amscd,amssymb}
\usepackage{color,epsfig}
\usepackage{xfrac}
\usepackage{latexsym}
\usepackage{mathrsfs}
\usepackage{ulem}
\usepackage[font=small,justification=centerlast,width=1.0\textwidth]{caption}

\newcommand{\fract}[2]{{\textstyle\frac{#1}{#2}}}

\begin{document}

\title{Collective Coordinates in One--Dimensional Soliton 
Models Revisited}

\author{I. Takyi, H. Weigel}

\affiliation{
Physics Department, Stellenbosch University,
Matieland 7602, South Africa}

\begin{abstract}
We compare numerical solutions to the full field equations to simplified 
approaches based on implementing three collective coordinates for kink--antikink 
interactions within the $\varphi^4$ and $\phi^6$ models in one time and 
one space dimensions. We particularly pursue the question whether the 
collective coordinate approximation substantiates the conjecture that 
vibrational modes are important for resonance structures to occur in 
kink--antikink scattering.
\end{abstract}

\maketitle

\section{Introduction}

The $\varphi^4$ model is a non--linear prototype extension of the Klein--Gordon 
theory in one time and one space dimensions which contains soliton solutions (more 
precisely solitary waves)~\cite{Ra82} that possess localized energy densities. 
Non--linearity gives rise to distinct vacuum 
solutions and the solitons connect different vacua at negative and positive spatial 
infinity. These solitons are called  {\sf kinks} and have a particle like behavior 
when subjected to external forces. Configurations obtained by spatial reflection 
are {\sf antikinks}. Similar soliton solutions are found in the {\sf sine--Gordon} 
and $\phi^6$ models. Non--integrability of the $\varphi^4$ and $\phi^6$ models makes 
them more interesting because of the enriched structure of solutions that correspond 
to kink antikink interactions. Localized soliton type solutions are not 
limited to these models. For example, lump--like configurations were obtained in 
modified $\varphi^4$ models~\cite{Avelar:2007hx} and kink--kink solutions in models 
with non--polynomial potentials~\cite{Mendonca:2015nka}. Solitons in the
$\phi^8$ model have been recently constructed in Ref.~\cite{Gani:2015cda}.

There is a wide range of applications for (anti)kink solutions in physics: In 
cosmology~\cite{Vachaspati:2006zz,Vilenkin:2000jqa} the kink solutions describe 
the fractal structure of domain walls~\cite{Anninos:1991un}; in condensed matter 
physics they mimic domain walls in ferromagnets~\cite{Ivanov:1992aa} and 
ferroelectrics~\cite{Trullinger:1976aa}. A remarkable feature of solitons is 
that their (classical) energy is inversely proportional to the coupling constant. 
Considering the number of colors in quantum chromodynamics as a hidden coupling 
constant~\cite{tHooft:1973jz} thus motivates to regard baryons as solitons in an 
effective meson theory~\cite{Witten:1979kh}. This picture of baryons has proven 
quite successful in describing many baryon properties~\cite{Weigel:2008zz}.
This approach has even been generalized to nuclei~\cite{Feist:2012ps}.
A general discussion of (topological) solitons can be obtained from 
Ref.~\cite{Manton:2004tk}.

Besides the zero mode associated with spontaneous breaking of translational 
invariance, the fluctuation 
spectrum about the $\varphi^4$ kink contains a further bound state, the so--called
shape mode. This shape mode possesses a number of interesting features. For
example, it can be indirectly excited by external forces~\cite{Quintero:2000aa}
and the frequency of wobbling kinks is correlated with the bound state energy
of the shape mode~\cite{Oxtoby:2009aa}. On the other hand the numerically observed
frequencies of oscillons (long living oscillations generated from an initial bump) 
are lower than the eigen--frequency of the shape mode~\cite{Fodor:2006zs}. This 
mode is a bound state in the background of the (anti)kink and thus represents an 
essential vibrational excitation of the (anti)kink. In kink--antikink scattering this 
mode might temporarily store energy and release it at a later time~\cite{Campbell:1983xu}. 
This process has been considered to be responsible for resonance phenomena in this 
scattering reaction: with energy stored in the shape mode, kink and antikink do 
not have enough energy to fully separate. Remarkably, such resonance solutions have 
also been observed in the $\phi^6$ model~\cite{Dorey:2011yw}. However, this model 
does not contain the shape mode. Of course, the interplay of translational and 
vibrational modes during the kink--antikink interaction is an interesting subject 
on its own that has been generalized to multiple kink interactions 
recently~\cite{Marjaneh:2016chl}.

The above survey is certainly incomplete but sufficient to demonstrate that 
there is a rich structure\footnote{See Ref.~\cite{Kivshar:1989aa} for an 
early review.} of solitons in non--linear low--dimensional models 
that can be identified by numerical simulations which are not too laborious. It 
is interesting and challenging to identify the dynamics behind these structures. 
A technique that has been frequently employed for this purpose is the introduction
of time dependent collective coordinates. They reduce the full field equations to
(coupled) ordinary differential equations. Though being an approximation relying on 
good guesses for appropriate collective coordinates, it assists to identify the 
relevant modes in case agreement between the solutions of the full and the reduced 
equations is obtained. Collective coordinates for the $\varphi^4$ model were
suggested in Ref.~\cite{Sugiyama:1979mi} already some time ago. Numerical 
calculations were only performed later on~\cite{Belova:1985fg,Anninos:1991un,Goodman:2005aa} 
and yielded remarkable agreement with the solutions to the full field equations, 
thereby stressing the relevance of the shape mode for resonance formation. 
Unfortunately, a typographical error in a formula from Ref.~\cite{Sugiyama:1979mi}
propagated into those numerical studies~\cite{Weigel:2013kwa}. Hence a second 
look at those studies is inevitable. Also, those studies included a subset of 
collective coordinates that is only appropriate for certain initial conditions. 
In addition a comprehensive collective coordinate study in the $\phi^6$ model 
is important to illuminate the origin of the resonance solutions within 
kink--antikink scattering. These issues will be the objectives of the present 
paper.

In section~\ref{models} we will introduce the models that we will consider. 
The collective coordinates will be defined in section~\ref{collcoord}. In 
section~\ref{numerics} we will present our numerical results. We will summarize 
with a short conclusion in section~\ref{concl}. An appendix details the 
calculation of coefficient functions in the collective coordinate approach.

\section{Models}
\label{models}

\noindent
The models that we consider are defined in one space and one time
dimensions. From the onset, we use dimensionless variables (corresponding
to $m=\sqrt{2}$ and $\lambda=2$ in Ref.~\cite{Sugiyama:1979mi}) so that 
the Lagrange densities are
\begin{equation}
\mathcal{L}_4=\frac{1}{2}\partial_\mu\varphi\partial^\mu\varphi
-\frac{1}{2}\left(\varphi^2-1\right)^2
\qquad {\rm and}\qquad 
\mathcal{L}_6=\frac{1}{2}\partial_\mu\phi\partial^\mu\phi
-\frac{1}{2}\phi^2\left(\phi^2-1\right)^2\,.
\label{eq:deflag}
\end{equation}
The respective field equations are partial differential
equations (PDE)
\begin{equation}
\ddot{\varphi}-\varphi^{\prime\prime}=2\varphi\left(1-\varphi^2\right)
\qquad {\rm and}\qquad
\ddot{\phi}-\phi^{\prime\prime}=-\phi\left(3\phi^4-4\phi^2+1\right)\,,
\label{eq:pde}
\end{equation}
that are straightforwardly obtained from the respective Lagrangians. 
In the above, dots and primes denote time and coordinate derivatives, 
respectively. There are two vacuum solutions in the $\varphi^4$ model, 
$\varphi_0=\pm1$ but three in the $\phi^6$ model, $\phi_0=\pm1$ and 
$\phi_0=0$. The PDE allow for static soliton solutions that connect
different vacuum solutions at spatial infinity. In the $\varphi^4$ model
these are the kink and antikink solutions
\begin{equation}
\varphi_{K,\overline{K}}(x)=\pm{\rm tanh}(x)\,,
\label{eq:kink4}
\end{equation}
that are related by spatial reflection $x\leftrightarrow-x$. In the $\phi^6$ 
model the soliton configurations that solve the field equations and connect the 
vacuum at $\phi_0=0$ with the vacuum $\phi_0=+1$ are
\begin{equation}
\phi_{K,\overline{K}}(x)=\left[1+{\rm exp}(\pm2x)\right]^{-\frac{1}{2}}\,.
\label{eq:kink6}
\end{equation}
Again these solutions are related by spatial reflection. In addition, the 
overall sign of $\phi_{K,\overline{K}}$ may be changed so that there are four 
different static solutions in the $\phi^6$ model. Time dependent solutions are 
straightforwardly constructed by a Lorentz boost: 
$x\longrightarrow\frac{x-vt}{\sqrt{1-v^2}}$, with constant velocity $v$.

There is an important difference between the two models that concerns the small 
amplitude fluctuations about the soliton solutions. While there are zero modes in 
both models that emerge because the static solutions break translational invariance 
spontaneously, the $\varphi^4$ model has an additional bound state solution, the 
so--called shape or breather mode~\cite{Ra82}. Parameterizing the field 
$\varphi(x,t)={\rm tanh}(x)+\eta(x,t)$ and linearizing the PDE in $\eta$, this shape
mode solution is found to be\footnote{The threshold is at $\omega=2$
for the dimensionless variables adopted here.}
\begin{equation}
\eta(x,t)={\rm e}^{-i\sqrt{3}t}\chi(x)
\qquad {\rm with} \qquad 
\chi(x)=\frac{{\rm sinh}(x)}{{\rm cosh}^2(x)}\,.
\label{eq:shape}
\end{equation}

The static solutions serve as initial conditions to investigate the kink--antikink 
system as a raw model for particle--antiparticle interactions. Initially a kink and
antikink are widely separated whilst moving towards each other. To be precise, in the 
$\varphi^4$ model the initial conditions read
\begin{align}
\varphi(x,0)&=\varphi_{\overline{K}}\left(\frac{x}{\sqrt{1-v^2}}-X_0\right)
+\varphi_K\left(\frac{x}{\sqrt{1-v^2}}+X_0\right)-1\,,\cr
\dot{\varphi}(x,0)&=\frac{v}{\sqrt{1-v^2}}\left[
\varphi^\prime_{\overline{K}}\left(\frac{x}{\sqrt{1-v^2}}-X_0\right)
-\varphi^\prime_K\left(\frac{x}{\sqrt{1-v^2}}+X_0\right) \right]\,,
\label{eq:in4}
\end{align}
where the primes denotes the derivative with respect to the argument. Here $X_0$ is a 
measure for the initial separation and $v$ is the relative kink and antikink velocity.

The situation is slightly more complicated in the $\phi^6$ model because two different
scenarios can be built up. First there is the kink--antikink configuration
\begin{align}
\phi_{K\overline{K}}(x,0)&=
\phi_{\overline{K}}\left(\frac{x}{\sqrt{1-v^2}}+X_0\right)
+\phi_K\left(\frac{x}{\sqrt{1-v^2}}-X_0\right)-1\,,\cr
\dot{\phi}_{K\overline{K}}(x,0)&=
\frac{-v}{\sqrt{1-v^2}}\left[
\phi^\prime_{\overline{K}}\left(\frac{x}{\sqrt{1-v^2}}+X_0\right)
-\phi^\prime_K\left(\frac{x}{\sqrt{1-v^2}}-X_0\right)\right]\,,
\label{eq:in6a}
\end{align}
and second the antikink--kink scenario
\begin{align}
\phi_{\overline{K}K}(x,0)&=
\phi_{\overline{K}}\left(\frac{x}{\sqrt{1-v^2}}-X_0\right)
+\phi_K\left(\frac{x}{\sqrt{1-v^2}}+X_0\right)\,,\cr
\dot{\phi}_{\overline{K}K}(x,0)&=
\frac{v}{\sqrt{1-v^2}}\left[
\phi^\prime_{\overline{K}}\left(\frac{x}{\sqrt{1-v^2}}-X_0\right)
-\phi^\prime_K\left(\frac{x}{\sqrt{1-v^2}}+X_0\right)\right]\,.
\label{eq:in6b}
\end{align}
Using the above initial conditions with large $X_0$, the numerical solutions of the
PDE~(\ref{eq:pde}) have been extensively discussed in the literature, both for the 
$\varphi^4$ 
\cite{Kudryavtsev:1975dj,Campbell:1983xu,Belova:1985fg,Anninos:1991un,Goodman:2005aa} 
and $\phi^6$ \cite{Dorey:2011yw} models. 
Especially in the $\varphi^4$ model the multifaceted structures, that emerge as the 
relative velocity $v$ is changed, have been widely explored. We will elaborate on these 
structures in section~\ref{numerics}. See also Ref. \cite{Weigel:2013kwa} for a 
comparative discussion for both models and a collection of further references.

\section{Collective coordinates}
\label{collcoord}

Time dependent collective coordinates have mainly been considered for the $\varphi^4$ 
model. Initially \cite{Sugiyama:1979mi} they were introduced to simplify the PDE to 
ordinary differential equations (ODE). Later, see {\it e.g.} \cite{Goodman:2005aa}
and references therein, they were utilized to explain the multiple bounce solutions 
within the kink--antikink collision that were earlier observed in the solutions
to the PDE. In this context the shape mode, Eq.~(\ref{eq:shape}), plays a decisive 
role. It has been conjectured that this mode is excited during the collision and 
that this excitation absorbs too much energy from the kink--antikink system to 
fully separate. Only when the shape mode releases this energy in phase 
with the dissociation of kink and antikink they possess enough energy to depart
to spatial infinity. Hence the amplitude of the shape mode and the kink--antikink
separation are important characteristics of the system and thus motivate to 
introduce corresponding collective coordinates via
\begin{equation}
\varphi_c(x,t)=\varphi_K(\xi_{+})+\varphi_{\overline{K}}(\xi_{-})-1
+\sqrt{\frac{3}{2}}\left[A(t)\chi(\xi_{-})+B(t)\chi(\xi_{+})\right]
\qquad {\rm where} \qquad
\xi_\pm=\xi_\pm(x,t)=\frac{x}{\sqrt{1-v^2}}\pm X(t)\,.
\label{eq:cc4}
\end{equation}
Obviously, $X(t)$ is the collective coordinate for the separation while 
$A(t)$ and $B(t)$ are the amplitudes of the shape modes in the background of 
the antikink and kink, respectively. This ansatz is substituted into the 
Lagrange density and, by spatial integration, a Lagrange function for
the collective coordinates is obtained. Generically it takes the form
\begin{align}
L_4(A,\dot{A},B,\dot{B},X,\dot{X})=a_1(X)\dot{X}^2-a_2(X)
&+a_3(X)\dot{A}^2-a_4(X)A^2+a_5(X)A+\ldots\cr
&+b_3(X)\dot{B}^2-b_4(X)B^2+b_5(X)B+\ldots-d_{10}(X)AB^3\,.
\label{eq:lc4}
\end{align}
The ellipsis refer to terms that involve other powers and products of the 
amplitudes $A$ and $B$ as derived from the Lagrangian, Eq.~(\ref{eq:deflag}). 
Unless otherwise noted, they are included in our calculation though omitted above 
for brevity only. The coefficient functions, $a_1(X),\ldots,d_{10}(X)$ are 
obtained as spatial integrals of the kink and/or antikink profile functions. 
For example, (see also the appendix),
$$
a_1(X)=\frac{1}{2}\int_{-\infty}^{\infty}dx\,\left[
\phi_K^\prime(\xi_{+})-\phi_{\overline{K}}^\prime(\xi_{-})\right]^2
=\frac{4}{3}\sqrt{1-v^2}\,
\left[1+\frac{6X{\rm coth}(2X)-3}{{\rm sinh}^2(2X)}\right]\,.
$$
Here we refrain from further listing those integrals in detail since they are 
documented in appendixes of Ref. \cite{Takyi:2015aa}. As indicated analytic 
expressions (as functions of $X$) can be obtained using analytic function 
theory~\cite{Sugiyama:1979mi,Moshir:1981ja,Campbell:1983xu}. Since we anyhow 
intend to perform numerical simulations, the numerical representation suits well. 
From the above Lagrange function a set of coupled second order ODE are derived 
that govern the time evolution of the collective coordinates. Schematically 
they are cast into the form
\begin{equation}
\begin{pmatrix}a_{11} & a_{12} & a_{13}\cr
a_{21} & a_{22} & a_{23}\cr
a_{31} & a_{32} & a_{33} \end{pmatrix}
\begin{pmatrix}\ddot{X} \cr \ddot{A} \cr \ddot{B}\end{pmatrix}
=\begin{pmatrix}f_1 \cr f_2 \cr f_3 \end{pmatrix}
\qquad {\Leftrightarrow} \qquad
\begin{pmatrix}\ddot{X} \cr \ddot{A} \cr \ddot{B}\end{pmatrix}
=\begin{pmatrix}a_{11} & a_{12} & a_{13}\cr
a_{21} & a_{22} & a_{23}\cr
a_{31} & a_{32} & a_{33} \end{pmatrix}^{-1}
\begin{pmatrix}f_1 \cr f_2 \cr f_3 \end{pmatrix}\,.
\label{eq:ode}
\end{equation}
The matrix elements, $a_{mn}$ and the right--hand sides, $f_n$ contain the
coefficients functions $a_i(X)$ etc. as well as factors of the collective 
coordinates themselves. For example,
$$
a_{11}=2\left[a_1(X)+a_6(X)A+a_8(X)A^2+b_6(X)B+b_8(X)B^2+d_2(X)AB\right]\,.
$$
The remaining lengthy formulas are reported in detail in Ref.~\cite{Takyi:2015aa}.
The ODE, Eq.~(\ref{eq:ode}) are to be solved for initial conditions resembling 
those for the PDE from section \ref{models}, {\it i.e.}, $X(0)=X_0$ and 
$\dot{X}(0)=\frac{-v}{\sqrt{1-v^2}}$. The amplitudes of the shape modes and their 
time derivatives all vanish initially.

The ansatz, Eq.~(\ref{eq:cc4}) was already proposed in Ref.~\cite{Sugiyama:1979mi}. 
However, numerical calculations have only be performed for reduced sets. For example,
the reduced system with $A(t)\equiv0$ and $B(t)\equiv0$ was solved in Ref.~\cite{Campbell:1983xu},
while the subsystem\footnote{For initial conditions that comply with this restriction
it remains a solution at all times.} $A(t)\equiv-B(t)$ was comprehensively investigated in
Ref.~\cite{Goodman:2005aa}. The latter restriction suffers from the obvious problem
that $\chi(\xi_{-})\sim\chi(\xi_{+})$ at zero separation $X(t)\sim0$ so that the 
amplitude $A(t)$ turns ill--defined. This is known as the null--vector 
problem~\cite{Caputo:1991cv}. This problem has been restated recently~\cite{Goodman:2015ina} 
in a different context but no rigorous soliton has been established so far. Furthermore, 
a typographical error in the formula for $a_5(X)$ from Ref.~\cite{Sugiyama:1979mi} has 
propagated through the literature making most of the numerical simulations obsolete.
We will give details in Sect.~\ref{varphi4}. Hence a re--analysis of these numerical 
approaches is essential.

The situation is slightly different in the $\phi^6$ model because there is no shape mode
in the fluctuation spectrum. Similar to the $\varphi^4$ model, however, multiple bounce 
solutions have also been observed from the PDE for the kink--antikink interaction in the $\phi^6$ 
model~\cite{Dorey:2011yw,Weigel:2013kwa}. As mentioned, the excitation of the shape mode 
during the kink--antikink interaction has been conjectured to cause the multiple bounce 
solutions in the $\varphi^4$ model. As a working hypothesis it thus seems very suggestive 
to include this degree of freedom as a representative of vibrational excitations
in the collective coordinate ansatz also for the $\phi^6$ model. 
We therefore write
\begin{align}
\overline{\phi}_{\rm cc}(x,t)&=
\phi_{K}(\xi_{-})+\phi_{\overline{K}}(\xi_{+})-1
+\sqrt{\fract{3}{2}}\left[A(t)\chi(\xi_{-})+B(t)\chi(\xi_{+})\right]\,,
\label{eq:cc6a} \\
\phi_{\rm cc}(x,t)&=
\phi_{K}(\xi_{+})+\phi_{\overline{K}}(\xi_{-})
+\sqrt{\fract{3}{2}}\left[A(t)\chi(\xi_{-})+B(t)\chi(\xi_{+})\right]\,,
\label{eq:cc6b} 
\end{align}
for the kink--antikink and antikink--kink systems, respectively. Again Lagrange functions
are computed for the two ans\"atze and second order ODE are obtained for the 
collective coordinates\footnote{Of course, we are treating three distinct models
defined by eqs.~(\ref{eq:cc4}), (\ref{eq:cc6a}) and~(\ref{eq:cc6b}). It is a matter
of convenience that no distinguishing symbols for the collective coordinates are 
introduced.} $X(t)$, $A(t)$ and $B(t)$. These equations are straightforwardly obtained 
but lengthy. The interested reader may extract them from Ref.~\cite{Takyi:2015aa}.
In Refs.~\cite{Goatham:2012tp,Gani:2014gxa} the calculation with $A(t)\equiv0$ and 
$B(t)\equiv0$ was performed. If that indeed was a sensible approximation, we should 
find that our extended parameterization in Eqs.~(\ref{eq:cc6a}) and~(\ref{eq:cc6b}) 
always yields negligible amplitudes of the vibrational shape mode.

\section{Numerical Results}
\label{numerics}

In this section we report the results from the numerical simulations of the 
differential equations discussed above. We solve the PDE equations, 
$\partial_t^2\phi(x,t)=\partial_x^2\phi(x,t)-V'(\phi(x,t))$,
as an initial value problem with the right hand side computed on a grid 
with typically 12001 equi--distant points for $x\in[-2X_0,2X_0]$. The second 
spatial derivatives are obtained from a five point formula that employs the 
actual position (coordinate argument on the left hand side of the PDE) and its 
two neighbors to the left and right. The PDE is then propagated in time with the help 
of an adaptive step size control. The initial configurations are taken from 
Eqs.~(\ref{eq:in4}), (\ref{eq:in6a}) and~(\ref{eq:in6b}). Though time consuming 
when so many points are implemented, this is a standard technique. 
In contrast to earlier studies of the ODE for the collective coordinates we use 
the numerical representation for the coefficient functions, $a_1(X)$, etc. 
that depend on the separation parameter $X$. Hence we do not have available these
coefficients for arbitrary values of $X$ needed in the numerical simulation. We 
therefore compute these coefficients for a large amount of values $X\in[-1.2X_0,1.2X_0]$
before attempting to solve the ODE. When integrating the ODE we then utilize a 
Laguerre interpolation to access the coefficients for the required $X$ as the ODE 
is propagated in time by an adaptive step size algorithm. To monitor the accuracy 
of numerically integrating the differential equations we verify that the total 
energy as obtained from the respective model Lagrangian is indeed time independent.

We want to compare solutions for the full field equations~(\ref{eq:pde}) to the
results from the collective coordinate approximation~(\ref{eq:ode}) with identical
initial conditions. This amounts to compare $X(t)$ to twice the distance between the 
kink and the antikink in the PDE calculation. We extract this from the expectation
value
\begin{equation}
\langle x \rangle_t=\frac{\int_0^\infty dx\, x\, \epsilon(t,x)}
{\int_0^\infty dx\,  \epsilon(t,x)}\,,
\label{eq:xn}
\end{equation}
where
\begin{equation}
\epsilon(t,x)=\frac{1}{2}\left[\ddot{\varphi}+\varphi^{\prime\prime}
+\left(\varphi^2-1\right)^2\right]
\qquad {\rm or} \qquad
\epsilon(t,x)=\frac{1}{2}\left[\ddot{\phi}+\phi^{\prime\prime}
+\left(\phi^2-1\right)^2\phi^2\right]
\label{eq:edens}
\end{equation}
are the energy densities of the respective models.
This procedure is based on the observation that the energy density
($\epsilon(t,-x)=\epsilon(t,x)$, when the initial field configuration is 
reflection symmetric) is characterized by two peaks that move in 
time. One peak signals the position of the kink on the negative half line, the 
other moves with the antikink on the positive half line. By restricting the
integration interval to $x\ge0$ in Eq.~(\ref{eq:xn}), the position of the 
antikink is identified.

\subsection{$\mathbf{\varphi^4}$ model}
\label{varphi4}

\begin{figure}
\centerline{
\epsfig{file=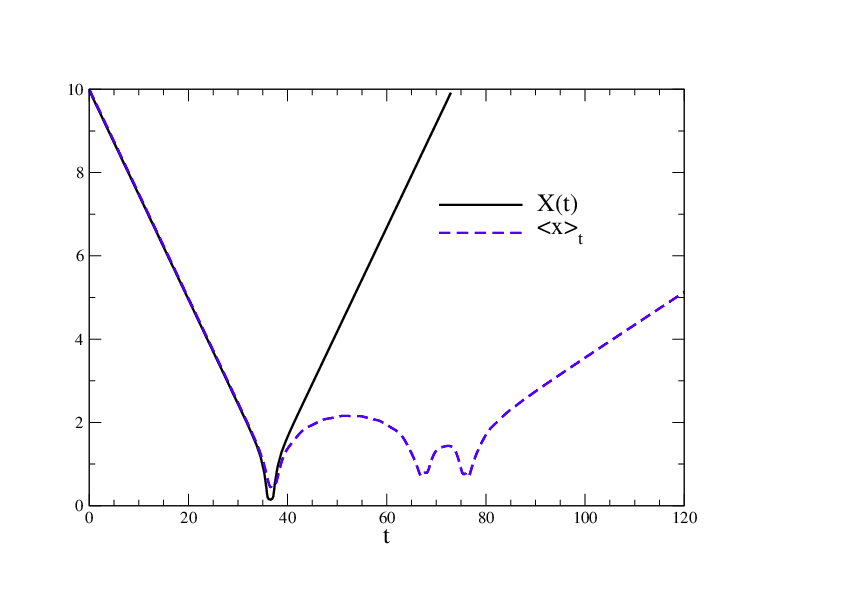,width=8cm,height=5cm}\hspace{1cm}
\epsfig{file=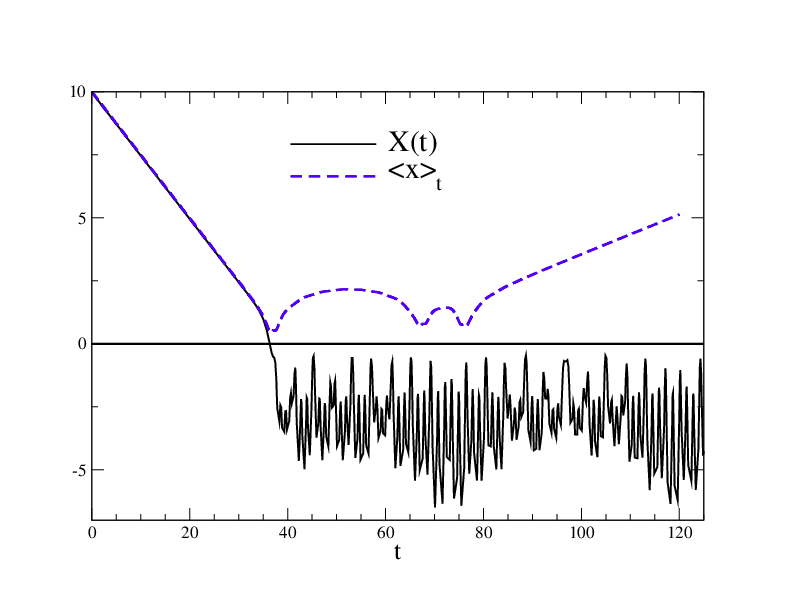,width=8cm,height=5cm}}
\caption{\label{fig:harm}(color online) Effect of correcting the linear coupling.
The full lines show the time dependence of the collective coordinate $X(t)$ and
the dashed lines picture the position of the antikink as extracted from
the PDE according to equation~(\ref{eq:xn}).  Left panel: calculation using $F(X)$ 
from Eq.~(\ref{eq:fxx}), right panel: corrected coupling from Eq.~(\ref{eq:a5corr}). 
Either case has $v=0.251$.\hfill\break}
\end{figure}

In a first step we explore the consequences of correcting the source term for 
the amplitude of the shape mode, the coefficients $a_5(X)$ and $b_5(X)$, in 
the numerical simulation of the $\varphi^4$ model. Because of a typographical 
error in Ref.~\cite{Sugiyama:1979mi} the expression (Here we list 
the formulas with $v=0$, for simplicity.)
\begin{equation}
F(X)=-3\pi\sqrt{\frac{3}{2}}\,{\rm tanh}^2(2X)\left[1-{\rm tanh}^2(2X)\right]
\label{eq:fxx}
\end{equation}
that entered previous simulations must be corrected to
\begin{equation}
a_5(X)=-3\pi\sqrt{\frac{3}{2}}\left[ 2-2\,{\rm tanh}^3(X)
-\frac{3}{{\rm cosh}^2(X)}+\frac{1}{{\rm cosh}^4(X)}\right]\,.
\label{eq:a5corr}
\end{equation}
We derive this corrected equation in the appendix. However, it is
intuitively clear that Eq.~(\ref{eq:fxx}) cannot be correct: the coefficient
$a_5$ is the amplitude of the modification of a background configuration. 
If that configuration is a solution to the field equations, the coefficient 
must vanish. It is easy to see that for $x=0$ the background
$\varphi_K(X)-\varphi_{\overline{K}}(X)-1=2{\rm tanh}(X)-1$ is not a 
vacuum configuration when $X\to-\infty$. Hence 
$\lim_{X\to-\infty}a_5(X)\stackrel{!}{\ne}0$. Obviously the expression
in Eq.~(\ref{eq:fxx}) violates this condition.\footnote{Besides the
convention regarding the arguments of the hyperbolic functions the essential
typo from Ref.~\cite{Sugiyama:1979mi} is the power of the ${\rm tanh}(X)$ 
term in Eq.~(\ref{eq:a5corr}). If that is decreased to 2, the expression in square 
brackets indeed simplifies to ${\rm tanh}^2(X)\left[{\rm tanh}^2(X)-1\right]$.}
Note that reasonable approximations to the full solutions were previously achieved
by using $F(X)$ for the linear coupling~\cite{Anninos:1991un,Goodman:2005aa}.
Additional simplifications were assumed in those simulations: (i) direct 
couplings between the shape modes at $\pm X$ and (ii) interactions involving
higher than quadratic powers of the shape were omitted. We reproduce the
result of such a calculation in the left panel of figure~\ref{fig:harm}.
There is indeed a similarity to the exact solution from the PDE. 
Once the correction for the linear coupling is implemented this similarity is completely 
lost. For the particular case of $v=0.251$ the kink--antikink system is trapped for an 
arbitrary long time during which it fluctuates around a negative value. In 
Ref.\cite{Weigel:2013kwa} similar fluctuations were reported for $v=0.2$, where,
however, the system separated after a long time. 

\begin{figure}
\centerline{
\epsfig{file=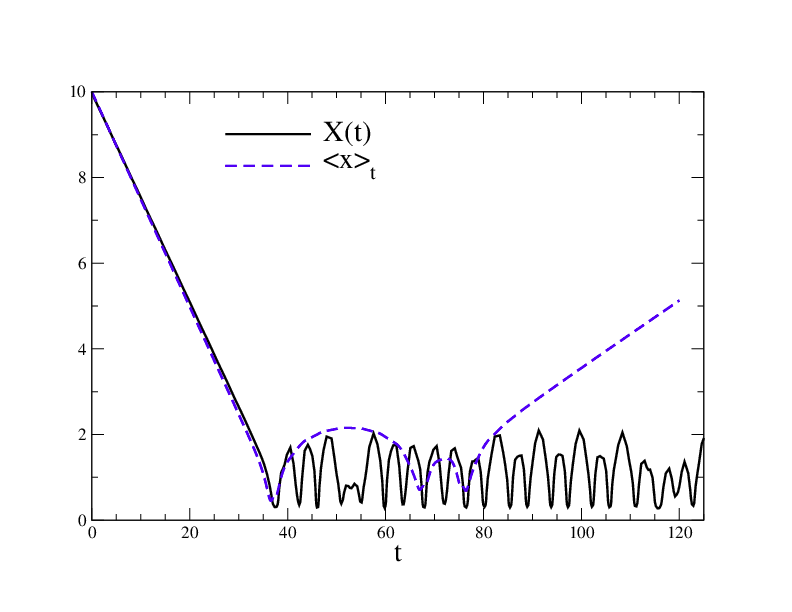,width=8cm,height=5cm}\hspace{1cm}
\epsfig{file=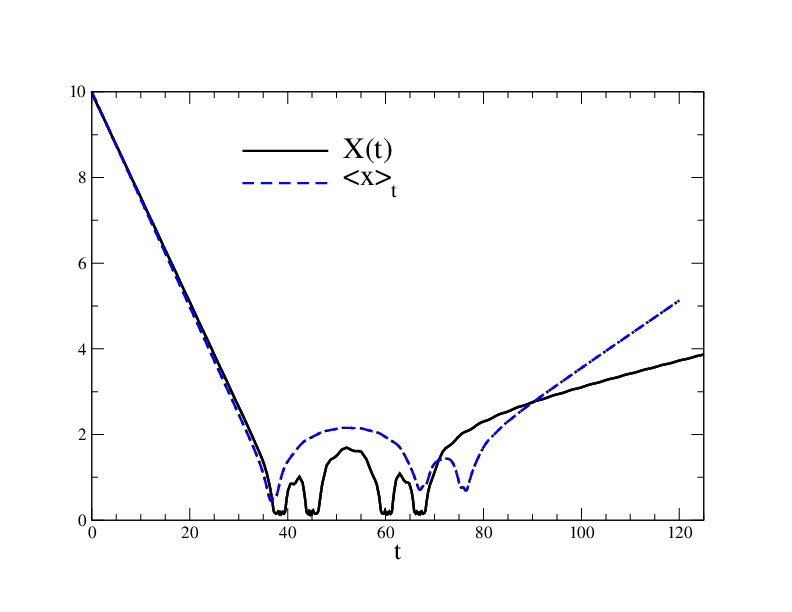,width=8cm,height=5cm}}
\caption{\label{fig:qdepend}(color online) Solutions to the ODE with $v=0.251$
and two different values $q=5$ (left panel) and $q=10$ (right panel) for the new
parameter in Eq.~(\ref{eq:cc4mod}).\hfill\break}
\end{figure}

The other approximations that were made previously are obviously possible explanations for
this mismatch between the ODE and PDE results. Omitting the higher order couplings
is also questionable in view of the results from Ref.~\cite{Abdelhady:2011tm}. A PDE calculation 
was performed for a wave packet moving towards a single $\varphi^4$ kink. Initially the wave
packet and the kink were well separated and after the interaction the phase shift was extracted 
from the wave packet. Agreement with the phase shift from the harmonic approximation required 
the amplitude of the fluctuations about it not to exceed 0.01, which is a small value in the 
present context.  (This does not invalidate formal results based on the harmonic approximation, 
like quantum corrections to kink properties that originate from the semi--classical
$\hbar$ expansion.) So it is suggestive to omit such approximations in the next step. 
However, then we face the problem that the matrix in Eq.~(\ref{eq:ode}) may become singular. 
Indeed this is the case.  As illuminated in the appendix, symmetry relations among the 
coefficient functions like, for example, $a_5(X)=-b_5(X)$ unfold that $A(t)=-B(t)$ is a 
possible solution to the ODE. Hence, configurations that initially obey this relation will 
do so for all $t$.  But then $A(t)$ and $B(t)$ are ill--defined when $X(t)=0$. This is the 
null--vector problem that causes the matrix from Eq.~(\ref{eq:ode}) to be singular. Of course, 
no such singularity is seen in the PDE. Furthermore in the PDE there is no (obvious) obstacle 
that prevents kink and antikink to penetrate. However, the collective coordinate 
parameterization is a not a solution to field equations for $X\to-\infty$ and 
$A(t)=B(t)=0$. These mismatches can be circumvented by a small modification of the 
collective coordinate parameterization. For the $\varphi^4$ model 
\begin{equation}
\varphi_c(x,t)=\varphi_K(\xi_{+})+\varphi_{\overline{K}}(\xi_{-})-{\rm tanh}(qX)
+\sqrt{\frac{3}{2}}\left[A(t)\chi(\xi_{-})+B(t)\chi(\xi_{+})\right]
\label{eq:cc4mod}
\end{equation}
is a possibility that introduces the new parameter $q>0$. This is an attempt to improve 
the collective coordinate ansatz and establish better agreement with the PDE results. 
Eventually it can also bypass the null--vector problem. Results for this
parameterization are shown in figure~\ref{fig:qdepend}. Obviously we observe only 
$X>0$ and see that the PDE results are better approximated as $q$ increases.
Even though $q>0$ was introduced to allow $X\to-\infty$ its introduction has the 
opposite effect. A large $q$, which corresponds to not modifying the parameterization 
as long as $X>0$, is needed to resemble the PDE results. This, however, induces 
large derivatives when $X\to0^{+}$ that absorb energy and in turn prevent the 
configuration to assume $X=0$.

\begin{figure}
\centerline{
\epsfig{file=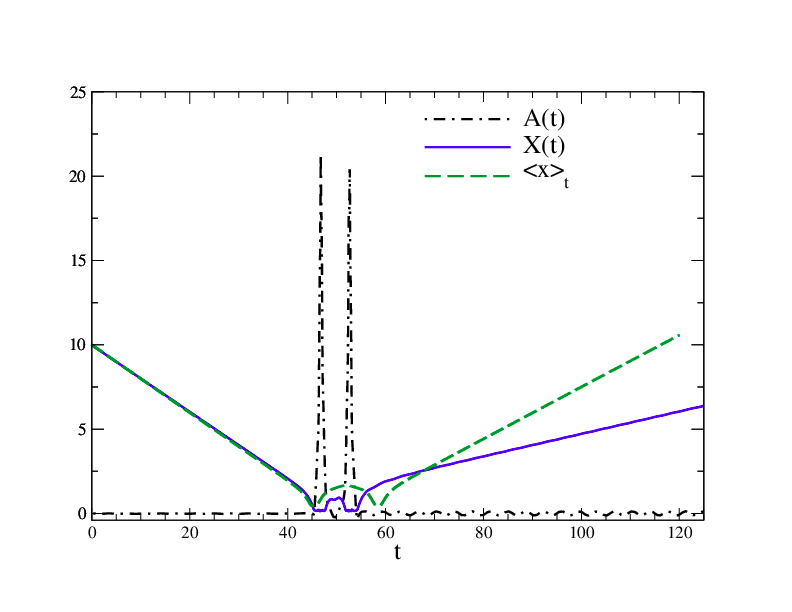,width=8cm,height=5cm}\hspace{1cm}
\epsfig{file=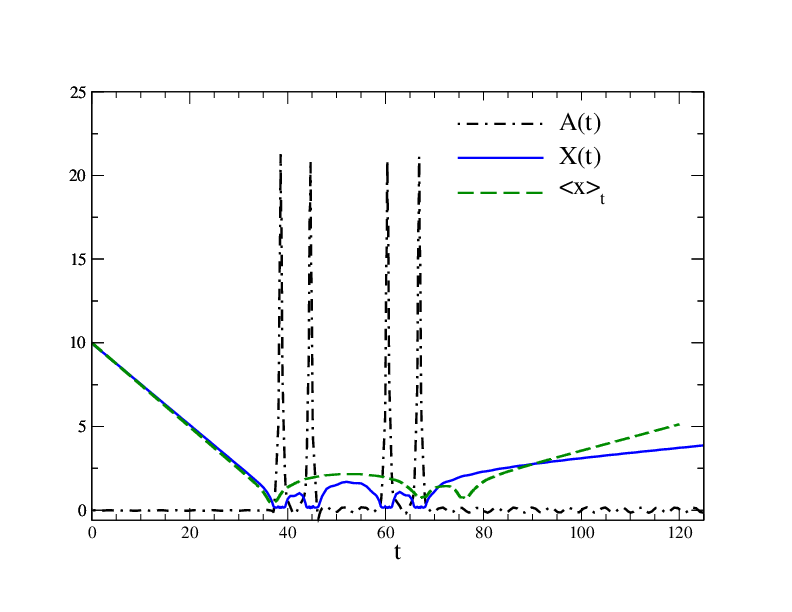,width=8cm,height=5cm}}
\caption{\label{fig:XA}(color online) Effect of shape mode in the $\varphi^4$ model.
ODE calculations with $q=10$ in Eq.~(\ref{eq:cc4mod}). Left panel: $v=0.201$, 
right panel: $v=0.251$.\hfill\break}
\end{figure}

In figure \ref{fig:XA} we compare the results for two different initial velocities
of the ODE to the corresponding solutions of the PDE with particular emphasis
on the amplitude of the shape mode. In each case the collective coordinate $X(t)$ has some 
qualitative similarity with the center of the energy density $\langle x\rangle_t$.
More importantly we observe that the amplitude of the shape mode is strongly enhanced as 
the kink--antikink system shrinks. This suggests that indeed a significant amount 
of energy is stored in that mode during the collision. Once the kink and antikink 
have separated this amplitude is essentially zero.

\subsection{$\mathbf{\phi^6}$ model}
\label{phi6}

As in the $\varphi^4$ model the original collective coordinate parameterizations,
Eqs.~(\ref{eq:cc6a}) and~(\ref{eq:cc6b}) describe vacuum configurations only 
for $X\ggg x$ but not for $X\lll x$. We therefore modify those parameterizations to
\begin{align}
\overline{\phi}_{\rm cc}(x,t)&=
\phi_{K}(\xi_{-})+\phi_{\overline{K}}(\xi_{+})
-\frac{1}{2}\left[{\rm tanh}(qX)+1\right]
+\sqrt{\fract{3}{2}}\left[A(t)\chi(\xi_{-})+B(t)\chi(\xi_{+})\right]\,,
\hspace{3cm}
\label{eq:cc6amod} \\
{\rm and}\hspace{3cm} & \cr
\phi_{\rm cc}(x,t)&=
\phi_{K}(\xi_{+})+\phi_{\overline{K}}(\xi_{-})
+\frac{3}{\sqrt{1+{\rm e}^{2qX}}}
+\sqrt{\fract{3}{2}}\left[A(t)\chi(\xi_{-})+B(t)\chi(\xi_{+})\right]\,.
\label{eq:cc6bmod} 
\end{align}
We have adjusted the additional constant $q\approx 10$ from numerical experiments on the 
qualitative agreement with the PDE results. 

\begin{figure}
\centerline{
\epsfig{file=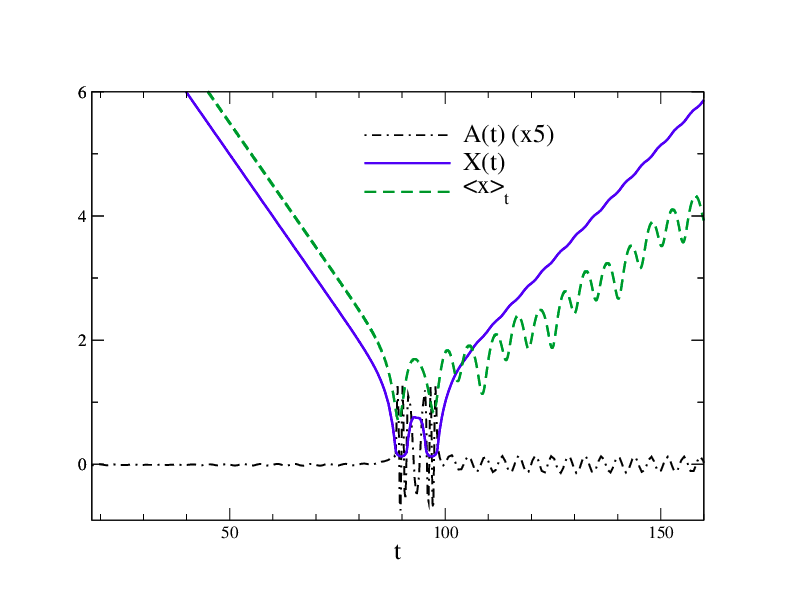,width=8cm,height=5cm}\hspace{1cm}
\epsfig{file=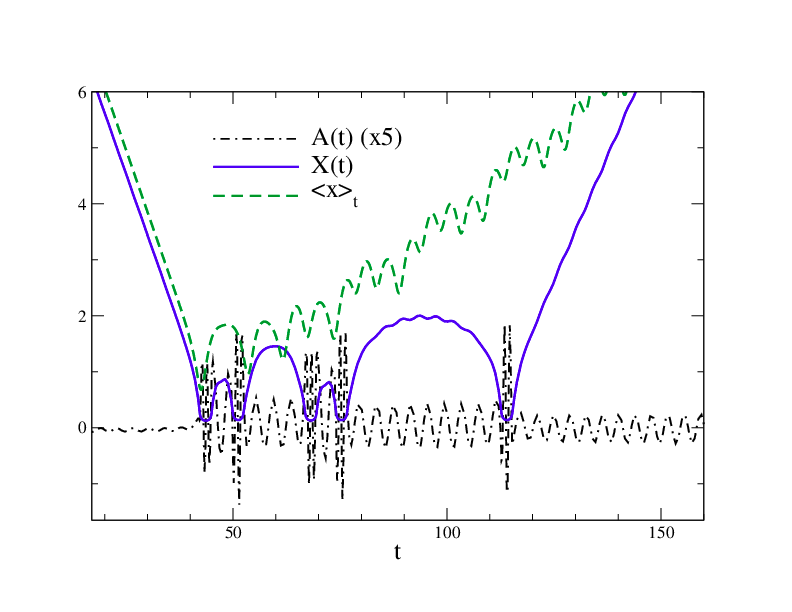,width=8cm,height=5cm}}
\caption{\label{fig:XAkak}(color online) Effect of shape mode in the $\phi^6$ model.
ODE calculations with $q=10$ in Eq.~(\ref{eq:cc6amod}). Left panel: $v=0.100$,
right panel: $v=0.221$.\hfill\break}
\end{figure}

In figure \ref{fig:XAkak} we show the results for the kink--antikink interaction. We observe
that the collective coordinate approximation reproduces the first resonance. However, at later
times we observe significant deviations from the exact PDE results. In particular, the collective
coordinate approximation does not reproduce the pronounced oscillations on top of the 
kink--antikink pair drifting apart with constant velocity.

\begin{figure}
\centerline{
\epsfig{file=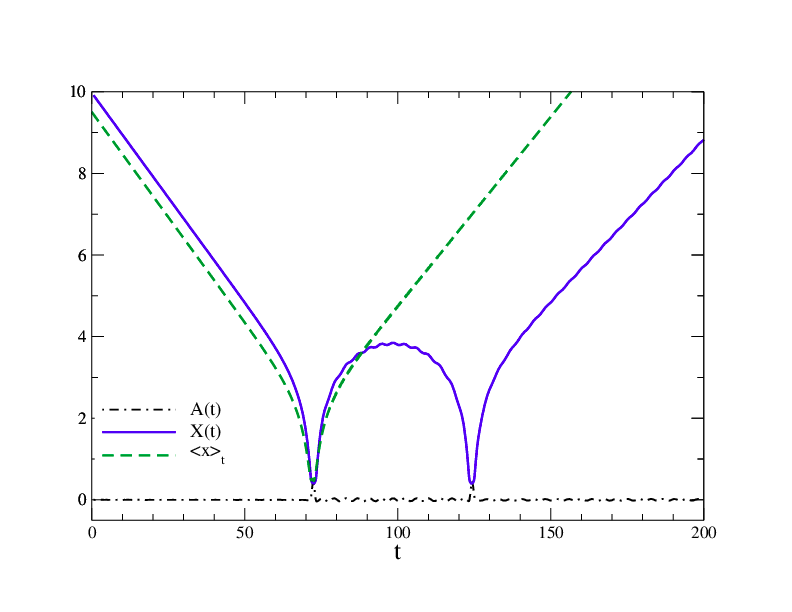,width=8cm,height=5cm}\hspace{1cm}
\epsfig{file=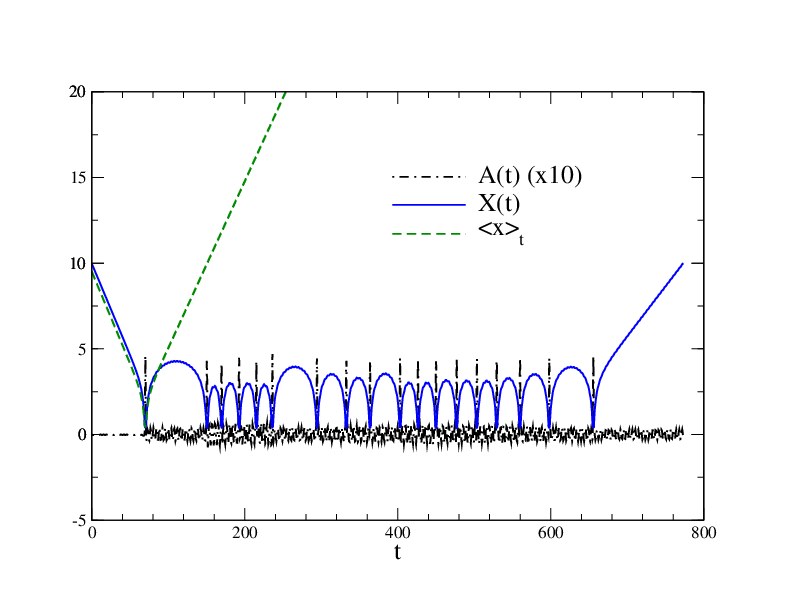,width=8cm,height=5cm}}
\caption{\label{fig:XAakk}(color online) Effect of shape mode in the $\phi^6$ model.
ODE calculations with $q=10$ in Eq.~(\ref{eq:cc6bmod}). Left panel: $v=0.103$,
right panel: $v=0.111$.\hfill\break}
\end{figure}

We show the results for the antikink--kink interaction arising from the initial condition 
of Eq.~(\ref{eq:cc6bmod}) in figure \ref{fig:XAakk}. In the two displayed cases the adopted 
initial velocity exceeds the critical velocity for bounces to occur in the PDE. In contrast, 
the ODE produces bounces and for $v=0.111$ there are many of them and the solution to 
the ODE is confined for a very long time.

As in the $\varphi^4$ model we see that indeed the shape mode gets excited as kink
and antikink get very close. However the corresponding amplitude $A(t)$ is not quite
as pronounced in the $\phi^6$ model. This suggests that other modes are also 
relevant for energy storage and resonance formation in the kink--antikink interaction 
of the $\phi^6$ model.

In figure \ref{fig:XAakknew} we finally consider the scenario for the antikink--kink 
interaction where the PDE produces bounces. The collective coordinate approximation does
so too, but the number of bounces differs in the two calculations. Even more
interestingly, their results from the PDE and ODE calculations are similar for moderate
times. But at even larger times the antikink--kink pair stays very closely together 
with the energy dwelling in the amplitude of the shape mode within the ODE solution.
\begin{table}
\begin{minipage}[c]{0.55\textwidth}
\centerline{
\epsfig{file=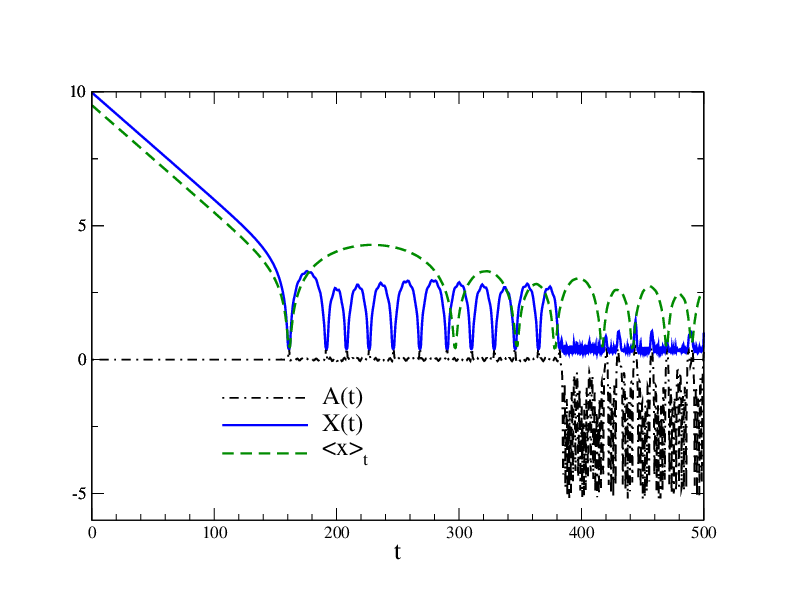,width=8cm,height=5cm}}
\captionsetup{width=1.0\columnwidth}
\captionof{figure}{\label{fig:XAakknew}(color online) Effect of shape mode in the $\phi^6$ model.
ODE calculations with $q=10$ in Eq.~(\ref{eq:cc6bmod}) for $v=0.040$.\hfill\break}
\end{minipage}
\hspace{1cm}
\begin{minipage}[c]{0.32\textwidth}
\renewcommand{\arraystretch}{1.5}
\centerline{\begin{tabular}{c|c|c}
system & PDE & ODE \cr
\hline
$\varphi^4$ & 0.26 & 0.4245 \cr
$\phi^6\,, K\overline{K}$ & 0.289 & 0.4424\cr
$\phi^6\,, \overline{K}K$ & 0.0457 & 0.1119
\end{tabular}}
\renewcommand{\arraystretch}{1.0}
\captionsetup{width=1.0\columnwidth}
\captionof{table}{\label{tab:vcr}Predictions for the critical velocities.\hfill\break}
\end{minipage}
\end{table}

In table \ref{tab:vcr} we list the predictions for the critical velocities above
which resonances cease to exist. We recognize that the collective coordinate
approximation reproduces the pattern but over estimates the exact results from 
the PDE.

\section{Conclusion}
\label{concl}

We have revisited the collective coordinate approximations to 
the field equations of the $\varphi^4$ and $\phi^6$ models in one time and
one space dimensions. Various arguments have motivated this investigation. 
First, there has been an inconsistency in the literature about the source 
term of the vibrational mode (represented by the shape mode) in the 
$\varphi^4$ model. Second, in that model this vibrational excitation has 
previously been considered as the driving force for resonance solutions 
in the kink--antikink interaction. However, that mode is not part of the 
small amplitude spectrum in the $\phi^6$ model which nevertheless contains 
resonance solutions. Third, identifying the amplitude of the shape modes for 
the kink and antikink leads to a singularity of the collective 
coordinate approximation when kink and antikink get arbitrarily close. To 
circumvent this so--called null--vector problem we have abandoned that identification. 
The structure of the equations of motion, however, revealed that this procedure
does not fully resolve this singularity for initial conditions that parameterize 
a pure kink--antikink system at large separation. Additional modifications of
the collective coordinate parameterization were necessary to achieve a non--singular 
description. These modifications were motivated by the search for a
better approximation to the exact PDE solutions and introduced a novel 
parameterization of the fields in terms of the collective coordinates.
With this new parameterization, the singular point was not part of the 
solution to the ODE and the null--vector problem did not occur.
Forth, many of the literature studies adopted the harmonic 
approximation for this amplitude (and other simplifications to avoid the 
null--vector problem). Since non--linearity is an essential feature of these 
models, it was inevitable to go beyond this approximation.

We have seen that the collective coordinate approximations resemble the 
solutions to the full field equations only moderately well. Some of the resonances
of the kink--antikink interactions are reproduced by solving the ODE of the
collective coordinate approximations, but these solutions typically produce too
many bounces. Also the ODE approach overestimates the relative initial velocities 
between kink and antikink above which no bounces occur in all scenarios. On the 
other hand we have seen that the amplitudes of fluctuations (modeled by the shape 
mode) about the kink--antikink system, are strongly enhanced during bounces, 
though it is significantly more pronounced in the $\varphi^4$ model than in the
$\phi^6$ model. This indeed suggests that energy is stored in these modes during 
the interaction. Unfortunately, the resemblance between the collective coordinate 
and the exact solutions is not good enough to turn that into the statement that 
the existence of a shape mode in the fluctuation spectrum is causal for the 
occurrence of multiple bounce solutions in kink--antikink scattering. Previous 
calculations with the distance between kink and antikink as the only collective 
coordinate ({\it e.g.} Ref.~\cite{Campbell:1983xu} for $\varphi^4$ and 
Ref.~\cite{Gani:2014gxa} for $\phi^6$ models, respectively) found acceptable 
agreement with the exact field equations. However, if that had been an appropriate 
approximation our generalization should not have yielded sizable amplitudes for 
the additional variables. 

\acknowledgments
This project is supported in part by the NRF (South Africa) under
grant~77454. The work of I.~Takyi was supported by a combined bursary from 
AIMS and Stellenbosch University. He appreciates current support from 
a STIAS fellowship.

\appendix

\section{Derivation of Eq.~(\ref{eq:a5corr})}

In this appendix we outline the derivation of the corrected source term
for the shape mode. There are linear coupling terms in the derivative
term $\partial_\mu \varphi\partial^\mu\varphi$ and in the potential 
$V(\varphi)=\fract{1}{2}\left(\varphi^2-1\right)^2$. We use the equation
of motion for $\varphi_{K,\overline{K}}$ and integrate by parts to 
write, with $V^\prime(\phi)=2\phi(\phi^2-1)$
\begin{align}
F(X)&=\int_{-\infty}^{\infty}dx\, \left\{V^\prime(\varphi_K(x+))
+V^\prime(\varphi_{\overline{K}}(x-X))-
V^\prime(\varphi_K(x+X)+\varphi_{\overline{K}}(x-X)-1)\right\}
\frac{{\rm sinh}(x+X)}{{\rm cosh}^2(x+X)}\cr
&=2\int_{-\infty}^{\infty}dx\,\Big\{
{\rm tanh}(x+X)\left[{\rm tanh}^2(x+X)-1\right]
-{\rm tanh}(x-X)\left[{\rm tanh}^2(x-X)-1\right]
\cr & \hspace{0.5cm}
-({\rm tanh}(x+X)-{\rm tanh}(x-X)-1)
\left[({\rm tanh}(x+X)-{\rm tanh}(x-X)-1)^2-1\right]\Big\}
\frac{{\rm sinh}(x+X)}{{\rm cosh}^2(x+X)}\,.
\label{eq:appF}
\end{align}
The boost factor $1/\sqrt{1-v^2}$ is straightforwardly included by
rescaling the integration variable.
The interesting terms are those in which $X$ appears with both 
signs. As an example we will work out 
\begin{equation}
I(X)=\int_{-\infty}^{\infty}dx\, {\rm tanh}^3(x-X)
\frac{{\rm sinh}(x+X)}{{\rm cosh}^2(x+X)}
\label{eq:appI1}
\end{equation}
in detail. As suggested in Ref.~\cite{Campbell:1983xu} we first
define the Fourier integral
\begin{equation}
I_k(X)=\int_{-\infty}^{\infty}dx\, 
{\rm e}^{ikx}\,
{\rm tanh}^3(x-X)
\frac{{\rm sinh}(x+X)}{{\rm cosh}^2(x+X)}
=\int_{-\infty}^{\infty}dz\, 
{\rm e}^{ikz}\,
\frac{{\rm sinh}^3(z-X)}{{\rm cosh}^3(z-X)}
\frac{{\rm sinh}(z+X)}{{\rm cosh}^2(z+X)}
\label{eq:appI2}
\end{equation}
and take the limit $I(X)=\lim_{k\to0}I_k(X)$. This Fourier integral
can be evaluated by analytic integration methods and noting that there
are two set of poles 
\begin{equation}
\mbox{1)~~second order poles at~} z=iy_n-X\hspace{2cm}
\mbox{2)~~third order poles at~} z=iy_n+X\,,
\nonumber\end{equation}
with $y_n=(2n+1)\fract{\pi}{2}$. We close the contour in the upper half plane 
so that $n=0,1,\ldots$ is relevant. To extract the residues
for the first set of poles we write $z=iy_n-X+\epsilon$ and expand
all functions under the integral to linear order in the small parameter
$\epsilon$
\begin{align}
{\rm cosh}(z+X)&\sim i(-1)^n\epsilon & 
{\rm cosh}(z-X)&\sim -i(-1)^n\left[s_2 -\epsilon c_2\right]\cr
{\rm sinh}(z+X)&\sim i(-1)^n & 
{\rm sinh}(z-X)&\sim -i(-1)^n\left[c_2 -\epsilon s_2\right]\cr
{\rm e}^{ikz}&\sim {\rm e}^{-iXk}{\rm e}^{-(2n+1)\pi k/2}
\left[1+ik\epsilon\right] \,, & 
\label{eq:expand1}
\end{align}
where we have abbreviated $c_2={\rm cosh}(2X)$ and $s_2={\rm sinh}(2X)$.
Then the integrand in Eq.~(\ref{eq:appI2}) expands as
\begin{align}
{\rm e}^{ikz}\,
\frac{{\rm sinh}^3(z-X)}{{\rm cosh}^3(z-X)}
\frac{{\rm sinh}(z+X)}{{\rm cosh}^2(z+X)}&=
i(-1)^n{\rm e}^{-iXk}{\rm e}^{-(2n+1)\pi k/2}
\left[1+ik\epsilon\right]
\frac{c_2^3-3\epsilon c_2^2s_2}{s_2^3-3\epsilon s_2^2c_2}
\frac{1}{\epsilon^2}+\mathcal{O}(\epsilon^0)\,.
\label{eq:expand2}
\end{align}
The relevant term involves $\frac{1}{\epsilon}$ and produces the
residue
\begin{equation}
R^{(1)}_n=(-1)^n{\rm e}^{-iXk}{\rm e}^{-(2n+1)\pi k/2}
\left[3i\frac{c_2^2}{s_2^4}-k\frac{c_2^3}{s_2^3}\right]\,.
\label{eq:rn1}
\end{equation}
Summing the geometric series 
$\sum_{n=0}^\infty (-1)^n{\rm e}^{-n\pi k}
=\left(1+{\rm e}^{-\pi k}\right)^{-1}$ yields
\begin{equation}
\sum_{n=0}^\infty R^{(1)}_n=
\frac{{\rm e}^{-iXk}}{2{\rm cosh}(\pi k/2)}
\left[3i\frac{c_2^2}{s_2^4}-k\frac{c_2^3}{s_2^3}\right]
\quad \longrightarrow \quad
\frac{3i}{2}\frac{c_2^2}{s_2^4}\qquad
{\rm as}\quad k\to0\,.
\label{eq:rn2}
\end{equation}
Since the geometric series resulted in an expression that is 
finite as $k\to0$, the $\mathcal{O}(k)$ parts in the expansion did 
not contribute to the final result. We now turn to the residues
of type 2) which are more cumbersome to evaluate because the
singularity is third order. Hence we must expand all functions
under the integral to one higher power when writing
$z=iy_n+X+\epsilon$
\begin{align}
{\rm cosh}(z+X)&\sim i(-1)^n\left[s_2+\epsilon c_2 
+\fract{1}{2}\epsilon^2s_2\right]&
{\rm cosh}(z-X)&\sim -i(-1)^n\epsilon\left[1+\fract{1}{6}\epsilon^2\right]\cr
{\rm sinh}(z+X)&\sim i(-1)^n\left[c_2+\epsilon s_2 
+\fract{1}{2}\epsilon^2c_2\right] & 
{\rm sinh}(z-X)&\sim -i(-1)^n\left[1+\fract{1}{2}\epsilon^2\right]\cr
{\rm e}^{ikz}&\sim {\rm e}^{iXk}{\rm e}^{-(2n+1)\pi k/2}
\left[1+ik\epsilon-\fract{1}{2}k^2\epsilon^2\right]\,. & 
\label{eq:expand3}
\end{align}
We recognize that the $n$ dependence is same as for the 
type 1) singularities so that it will be sufficient to only keep 
terms that do not vanish in the limit $k\to0$ when expanding 
the function under the integral in Eq.~(\ref{eq:appI2})
\begin{align}
{\rm e}^{ikz}\,
\frac{{\rm sinh}^3(z-X)}{{\rm cosh}^3(z-X)}
\frac{{\rm sinh}(z+X)}{{\rm cosh}^2(z+X)}&=
-i(-1)^n{\rm e}^{iXk}{\rm e}^{-(2n+1)\pi k/2}
\left\{
\frac{\left(1+\frac{\epsilon^2}{2}\right)^3}
{\left(1+\frac{\epsilon^2}{6}\right)^3}
\frac{c_2+\epsilon s_2 +\fract{\epsilon^2}{2}c_2}
{\left[s_2+\epsilon c_2 +\fract{\epsilon^2}{2}s_2\right]^2}
\frac{1}{\epsilon^3}+\mathcal{O}(k)\right\}
+\mathcal{O}(\epsilon^0)\,.
\label{eq:expand4}
\end{align}
We read off the residue from the term with $\frac{1}{\epsilon}$:
\begin{align}
R^{(2)}_n&=-i(-1)^n{\rm e}^{iXk}{\rm e}^{-(2n+1)\pi k/2}
\left\{\frac{1}{s_2^4}\left[-\fract{1}{2}c_2s_2^2+3c_2^3-c_ss_2^2\right]
+\mathcal{O}(k)\right\}\cr
&=-i(-1)^n{\rm e}^{iXk}{\rm e}^{-(2n+1)\pi k/2}
\left\{\frac{3}{2}\frac{c_2}{s_s^4}\left[c_2^2+1\right]
+\mathcal{O}(k)\right\}\,.
\label{eq:expand5}
\end{align}
We can now sum the residues and take the limit $k\to0$
\begin{equation}
\sum_{n=0}^\infty R^{(2)}_n=
-i\frac{{\rm e}^{-iXk}}{2{\rm cosh}(\pi k/2)}
\left\{\frac{3}{2}\frac{c_2}{s_2^4}\left[c_2^2+1\right]
+\mathcal{O}(k)\right\}
\quad \longrightarrow \quad
-\frac{3i}{4}\frac{c_2}{s_2^4}\left[c_2^2+1\right]\,.
\label{eq:rn3}
\end{equation}
Finally we multiply the sum of Eqs.~(\ref{eq:rn2}) and~(\ref{eq:rn3})
by $2\pi i$ and obtain 
\begin{equation}
I(X)=\frac{3\pi}{2}\frac{1}{s_2^4}\left[c_2^3+c_2-2c_2^2\right]\,.
\label{eq:final_integral}
\end{equation}
Using the addition theorems $s_2=2{\rm sinh}(a){\rm cosh}(a)$ and
$c_2=2{\rm cosh}^2(a)-1=2{\rm sinh}^2(a)+1$ gives a further simplification 
that we list below together with other relevant integrals that are 
obtained by the same techniques:
\begin{align}
\int_{-\infty}^{\infty}dx\, {\rm tanh}^3(x-X)
\frac{{\rm sinh}(x+X)}{{\rm cosh}^2(x+X)}
&=\frac{3\pi}{8}\frac{2{\rm cosh}^2X-1}{{\rm cosh}^4X}
\cr
\int_{-\infty}^{\infty}dx\, {\rm tanh}^2(x-X)
{\rm tanh}(x+X)\frac{{\rm sinh}(x+X)}{{\rm cosh}^2(x+X)}
&=\frac{\pi}{8}\frac{4{\rm cosh}^4X-4{\rm cosh}^2X+3}{{\rm cosh}^4X}
\cr
\int_{-\infty}^{\infty}dx\, {\rm tanh}(x-X)
{\rm tanh}^2(x+X)\frac{{\rm sinh}(x+X)}{{\rm cosh}^2(x+X)}
&=\frac{\pi}{8}\frac{4{\rm cosh}^2X-1}{{\rm cosh}^4X}
\cr
\int_{-\infty}^{\infty}dx\, {\rm tanh}(x-X)
{\rm tanh}(x+X)\frac{{\rm sinh}(x+X)}{{\rm cosh}^2(x+X)}
&=\frac{\pi}{4}\frac{{\rm tanh}X}{{\rm cosh}^2X}
\left[1-2{\rm cosh}^2X\right]
\cr
\int_{-\infty}^{\infty}dx\, {\rm tanh}(x-X)^2
\frac{{\rm sinh}(x+X)}{{\rm cosh}^2(x+X)}
&=-\frac{\pi}{2}\frac{{\rm tanh}X}{{\rm cosh}^2X}
\cr
\int_{-\infty}^{\infty}dx\, {\rm tanh}(x-X)
\frac{{\rm sinh}(x+X)}{{\rm cosh}^2(x+X)}
&=\frac{\pi}{2}\frac{1}{{\rm cosh}^2X}\,.
\label{eq:list}
\end{align}
We have verified these integrals by numerical simulation.
Some of the integrals are even in $X$ while others are odd. Hence
\begin{equation}
F(X)=-3\pi\left[2-2{\rm tanh}^3X-3\frac{1}{{\rm cosh}^2X}
+\frac{1}{{\rm cosh}^4X}\right]
\label{eq:fx}
\end{equation}
does not have a specified transformation property under $X\leftrightarrow-X$,
in contrast to the erroneous literature formula quoted in Eq.~(\ref{eq:fxx}).
Terms under the integral in Eq.~(\ref{eq:appF}) that only have $x+X$ as 
arguments of the hyperbolic functions were obtained from the above list 
by setting $X=0$. Multiplying with the normalization of the shape mode 
($\sqrt{\frac{3}{2}}$), gives the coefficient $a_5$ provided in
Eq.~(\ref{eq:a5corr}). 

To compute the coefficient $b_5$ we need to replace 
\begin{equation}
\frac{{\rm sinh}(x+X)}{{\rm cosh}^2(x+X)}
\quad \longrightarrow \quad
\frac{{\rm sinh}(x-X)}{{\rm cosh}^2(x-X)}
=-\frac{{\rm sinh}(-x+X)}{{\rm cosh}^2(-x+X)}\,.
\end{equation}
in Eq.~(\ref{eq:appF}). Then $b_5=-a_5$ because the term in curly brackets
is invariant under $x\to-x$. This is one example for a general property
of the collective coordinate Lagrangians, Eq.~(\ref{eq:lc4}) and its 
$\phi^6$ pendant; they are invariant under $A\leftrightarrow-B$.

For completeness we also outline the result for $a_1$ listed in section III. 
It is interesting because the limit $k\to0$ works differently from the above.
Since $\frac{d}{dx}{\rm tanh}(x)=1/{\rm cosh}^2(x)$ the separation dependent
part of $a_1$ requires the $k\to0$ limit of
\begin{equation}
\int_{-\infty}^\infty dz \,\frac{{\rm e}^{ikz}}{{\rm cosh}^2(x-X){\rm cosh}^2(x+X)}\,.
\label{eq:a1a}
\end{equation}
The singularities along the lines $\mbox{\sf Re}(z)=\pm X$ are second order 
and it suffices to consider Eqs.~(\ref{eq:expand1}) and~(\ref{eq:expand3}) to 
linear order in $\epsilon$. The residues are 
\begin{equation}
{\rm e}^{\pm iXk}{\rm e}^{-(2n+1)\pi k/2}\frac{1}{s_2^2}
\left[ik\mp2\frac{c_2}{s_2}\right]
\qquad {\rm for} \qquad
z=iy_n\pm X\,.
\label{eq:a1b}
\end{equation}
Summing all contributions yields
\begin{equation}
\frac{1}{2{\rm sinh}(\pi k)}\left[
2\frac{c_2}{s_2}\left({\rm e}^{-iXk}-{\rm e}^{+iXk}\right)
+ik\left({\rm e}^{-iXk}+{\rm e}^{+iXk}\right)\right]
\quad \longrightarrow\quad 
\frac{-i}{\pi s_2^2}\left(2X\frac{c_2}{s_2}-1\right)
\qquad {\rm as} \quad k\to0\,.
\label{eq:a1c}
\end{equation}
This is the separation dependent part of $a_1$.

\bibliographystyle{apsrev}

\end{document}